\documentclass
[preprint,showpacs,preprintnumbers,amsmath,amssymb,floatfix]{revtex4}

\usepackage{epsf}

\usepackage{amsmath}

\newcommand{\bVec}[1]{\mbox{\boldmath$#1$}}

\begin{document}

\preprint{}

\title{Polarization observables in exclusive kaon photoproduction
 on the deuteron}

\author{K. Miyagawa}
\affiliation{Department of Applied Physics, Okayama University
of Science\\
1-1 Ridai-cho, Okayama 700, Japan}

\author{T. Mart}
\affiliation{Departemen Fisika, FMIPA, Universitas Indonesia, Depok
 16424, Indonesia}

\author{C. Bennhold}
\affiliation{Center for Nuclear Studies, Department of Physics, 
The George Washington 
  University, Washington, D.C. 20052, USA}

\author{W. Gl\"ockle}
\affiliation{Institut f\"ur Theoretische Physik II,
 Ruhr-Universit\"at Bochum, 
  D-44780 Bochum, Germany}

\begin{abstract}
Single and double polarization observables for kaon photoproduction
on the deuteron are studied theoretically with modern hyperon-nucleon
forces. 
The kinematical region of the  kaon scattered forward 
with large momentum is thoroughly investigated
where either the quasifree scattering leading to the kaon and hyperon
or the final-state interaction between hyperon and nucleon are expected.
The quasifree scatterings show characteristic peaks in the inclusive
cross sections. 
The final-state interaction effects are significant,
especially around the $\Lambda N$ and $\Sigma N$ thresholds.
The double polarization $C_z$ is found to be sensitive to
the final-state interaction effects.
Precise data would help to clarify the property of 
the $\Lambda N$-$\Sigma N$ interaction and also
help to extract the information on the elementary
amplitude from the quasifree scattering region.   
\end{abstract}


\maketitle

\section{Introduction}
\label{intro} 

The hyperon-nucleon ($YN$) interaction has played a key 
role in hypernuclear physics. It is known as the basic force which
binds $\Lambda$ and $\Sigma$ hyperons in hypernuclei. Even in a
conventional nuclear system, understanding of the $YN$ interaction is 
indisputably important if one introduces the strangeness degrees of
 freedom
in the nucleon-nucleon ($NN$) interaction, in order to extend 
the baryon-baryon interaction to a more unified picture demanded
by SU(3) symmetry.

While performing $YN$ scattering experiments is a daunting task,
hyperon photoproduction on the deuteron appears to be a promising
process for investigating the available $YN$ interaction models. 
This is made possible by choosing some photoproduction kinematics 
where the outgoing baryons have small relative momenta and, 
therefore, the $YN$ interaction is effectively strong. 
Several previous studies have been done in inclusive and exclusive kaon 
photoproduction on the deuteron \cite{adelseck89} based on simple
$YN$ forces. Nevertheless, all calculations indicated that significant 
$YN$ final state
interaction effects are present near the production thresholds.

Another motivation of performing experiments on the deuteron is
due to the lack of the neutron target. Experiments for kaon
photoproduction on the proton have just been finished in ELSA, JLab, 
and SPRING8, resulting in  a large number of unprecedented precise data 
that certainly need more sophisticated analyses. However, previous
 studies
\cite{mart1995,saghai1996} had indicated that neutron channels can 
provide a more stringent constraint to elementary models that
can successfully explain the process on the proton. Thus,
kaon photoproduction on the deuteron can serve as a means
for testing the available elementary models. To this end,
a simple calculation has proven that in most kinematic situations 
the elementary cross section on the neutron can be reliably 
extracted from the deuteron cross section \cite{Li:1992tz}.
A related experiment has recently been done at LNS-Tohoku
by measuring the $K^0 YN$ final states cross section and the 
collected data are being analyzed \cite{tohoku2004}.

In the previous paper \cite{yamamura99}, we have investigated the 
inclusive $K^+$ and exclusive $K^+Y$ photoproduction on the deuteron
by using modern hyperon-nucleon forces and a certain 
kaon photoproduction operator. Sizable
 effects of the hyperon-nucleon final state interaction were
found  near the $\Lambda N$ and $\Sigma N$ thresholds 
in the inclusive reaction.
Angular distributions for the exclusive process 
showed clear $YN$ final state interaction effects in certain
kinematic regions. We pointed out that 
precise data especially for the inclusive process around the
$\Sigma N$ threshold would help to clarify the 
strength and property of the
$\Lambda N$-$\Sigma N$ interaction. 

We notice that a  fairly elaborate calculation focusing on rescattering
 contributions
to the kaon photoproduction on the deuteron has recently appeared 
\cite{maxwell2004}. In order to emphasize the influence of rescattering 
terms in the differential cross sections, the calculation, however, has 
been performed in a certain kinematics where final state interactions 
among the hyperons and the nucleon have negligible effects.
A more detailed calculation, which includes
the $YN$ and $KN$ rescattering terms, as well as a two-body
 photoproduction
contribution in terms of a pion mediated process
 $\gamma d\to \pi NN\to KYN$, 
has also been performed \cite{salam2004}. It is found that the latter
 gives
the dominant contribution beyond the impulse approximation for backward
kaon angles. In the forward directions ($\theta_K \lesssim 30^\circ$)
 the effects
turn out to be small.

In this paper we theoretically  examine in detail a number of
 polarization
observables that are immediately accessible experimentally due to
the high intensity continuous wave electron beams available nowadays 
at JLab, ELSA, GRAAL, and SPRING8, and due to the hyperon recoil 
polarization, which can be directly obtained from the hyperon decay. 
This work is also greatly motivated
by the  Hall A and Hall B data recently taken which are currently 
being analyzed \cite{mecking89,reinhold91}.

For the convenience of the reader 
we briefly describe the latest update of the elementary photoproduction
operator in Sec.~II.
Section~III describes
the evaluation of the nuclear matrix element and the
definition of various polarization observables.
Our numerical results for the various $YN$ forces are
extensively given in Sec.~IV and we give the conclusion in Sec.~V.

\section{The elementary production operator}

Models for kaon photoproduction on the nucleon have been proliferating
rapidly in recent years, due mostly to the continuously improving data 
base
from experimental facilities like ELSA, GRAAL, and Jlab.  Most of these
descriptions employ an effective Lagrangian approach at tree 
level \cite{mart1999,mart2000,janssen2002,saghai1996} where the results 
from coupled-channel calculations \cite{penner2002} are taken into account.
  While the details
of the model ingredients vary in the different studies, all models provide
 a good
parameterization of the elementary  kaon photoproduction data and are
 thus adequate for use in the nuclear environment.  

Recently, we note that two new versions of data have just been published 
by the SAPHIR and the CLAS
collaborations \cite{Glander:2003jw,Bradford:2005pt}. 
Unfortunately, as reported in Ref.\,\cite{Bradford:2005pt}, 
the CLAS data show substantial and systematic discrepancies with 
the SAPHIR ones. Thus, as a consequence, an attempt to fit the model to both 
data simultaneously would be meaningless. While waiting for a new method
to reconcile this problem, and for the sake of
comparison, we will in this study use the same operator \cite{fxlee2001}
 we have
used in our previous paper \cite{yamamura99}.  
The operator consists of gauge-invariant
background and resonances terms. The background terms include
 the standard $s$-,
$u$-, and $t$-channel contributions along with a contact term required to
 restore
gauge invariance after hadronic form factors have been introduced.
The resonance part consists of three nucleon
resonances that have been found in the coupled-channels approach to decay
 into the
$K\Lambda$ channel, the $S_{11}$(1650), $P_{11}$(1710), and
 $P_{13}(1720)$. 
For the $K \Sigma$ production operator further contributions from 
the $S_{31}$(1900) 
and $P_{31}$(1910) $\Delta$- resonances were added. In addition, we added
 the
$D_{13}(1900)$ state \cite{mart1999} found to be important in the
 description 
of SAPHIR data \cite{Glander:2003jw,saphir}. 

In order to implement the operator in a nuclear matrix element, we
rewrite the amplitude in an arbitrary frame

\begin{eqnarray}
t_{\gamma K}&=&\left(\frac{E_N+m_N}{2m_N}\right)^{\frac{1}{2}}
\left(\frac{E_Y+m_Y}{2m_Y}\right)^{\frac{1}{2}}\sqrt{\frac{m_Y}{E_Y}}
\sqrt{\frac{m_N}{E_N}}\times\nonumber\\
&&\nonumber\\
&&\left[\ 
   {\cal F}_1 \,   \bVec{\sigma}\cdot\bVec{\epsilon}
+ {\cal F}_4 \, \bVec{\sigma}\cdot\bVec{p}_\gamma\,\bVec{p}_N\cdot
                                                           \bVec{\epsilon}
+ {\cal F}_5 \, \bVec{\sigma}\cdot\bVec{p}_\gamma\,\bVec{p}_Y\cdot
                                                          \bVec{\epsilon}
+ {\cal F}_8 \, \bVec{\sigma}\cdot\bVec{p}_N\,\bVec{p}_N\cdot\bVec{\epsilon}
+ {\cal F}_9  \,\bVec{\sigma}\cdot\bVec{p}_N\,\bVec{p}_Y\cdot\bVec{\epsilon}
\right.
\nonumber\\
&&
+~ {\cal F}_{12}\, \bVec{\sigma}\cdot\bVec{p}_Y\,\bVec{p}_N\cdot
\bVec{\epsilon}
+ {\cal F}_{13}\, \bVec{\sigma}\cdot\bVec{p}_Y\,\bVec{p}_Y\cdot\bVec{\epsilon}
+ {\cal F}_{14}\, \bVec{\sigma}\cdot\bVec{\epsilon}\,\bVec{\sigma}\cdot
                                \bVec{p}_\gamma\bVec{\sigma}\cdot\bVec{p}_N
\nonumber\\
&&
+~ {\cal F}_{15}\, \bVec{\sigma}\cdot\bVec{p}_Y\,\bVec{\sigma}\cdot
                                                         \bVec{\epsilon}
                \bVec{\sigma}\cdot\bVec{p}_\gamma 
+ {\cal F}_{16}\, \bVec{\sigma}\cdot\bVec{p}_Y\,\bVec{\sigma}\cdot
                                                         \bVec{\epsilon}
                \bVec{\sigma}\cdot\bVec{p}_N
+ {\cal F}_{19}\, \bVec{\sigma}\cdot\bVec{p}_Y\,\bVec{\sigma}\cdot
                                                        \bVec{p}_\gamma
                \bVec{\sigma}\cdot\bVec{p}_N\,\bVec{p}_N\cdot\bVec{\epsilon}
\nonumber \\
&&
+ \left. {\cal F}_{20}\, \bVec{\sigma}\cdot\bVec{p}_Y\,\bVec{\sigma}\cdot
                                                         \bVec{p}_\gamma
                \bVec{\sigma}\cdot\bVec{p}_N\,\bVec{p}_Y\cdot\bVec{\epsilon}
\ \right] ~,
\label{op1} 
\end{eqnarray}
with $m_N$ and $m_Y$ are the nucleon and hyperon masses, $E_N$ and $E_Y$
their energies, $\bVec{p}_\gamma$, $\bVec{p}_N$ and $\bVec{p}_Y$ the photon,
nucleon and hyperon momenta and $\bVec{\epsilon}$ the photon polarization.
The amplitudes ${\cal F}_i$ are given in terms of kinematical quantities
and amplitudes  $A_i$  which are related to the various tree diagrams. 
The rather lengthy expressions  of ${\cal F}_i$ and  $A_i$
can be found in \cite{mart_thesis}.
The operator can be rewritten in the form of 
\begin{equation} 
t_{\gamma K} = i\left(\,L\,+\,i\bVec{\sigma}\cdot\bVec{K}\,\right) ~,
\label{op2} 
\end{equation}
which is convenient for an application in the deuteron frame. However,
in Eq.~(\ref{op2}) both spin-non-flip and spin-flip amplitudes $L$ and 
$\bVec{K}$ depend on the photon polarization $\bVec{\epsilon}$ which
 complicates
the calculation on the deuteron. To avoid this, we rewrite Eq.~(\ref{op1})
in the form of
\begin{eqnarray} 
t_{\gamma K} &=& i\,(1,i\sigma_x,i\sigma_y,i\sigma_z) \,
\left( \begin{array}{ccc}
{\cal L}_x & {\cal L}_y & {\cal L}_z \\ {\cal K}_{xx} & {\cal K}_{xy} &
 {\cal K}_{xz} \\ 
{\cal K}_{yx} & {\cal K}_{yy} & {\cal K}_{yz} \\ 
{\cal K}_{zx} & {\cal K}_{zy} & {\cal K}_{zz} \end{array}\right) ~
\left( \begin{array}{c}
\epsilon_x \\ \epsilon_y \\ \epsilon_z  \end{array}\right) ~.
\label{op3} 
\end{eqnarray}
Therefore, the elementary operator given by the $4\times 3$ matrix in 
Eq.~(\ref{op3}) becomes completely independent from the frame in which 
$\bVec{\epsilon}$ and $\bVec{\sigma}$ are defined. The expressions for
the elements of this matrix read
\begin{eqnarray} 
\bVec{\cal L} &=& N \left[{\cal F}_{14}\,\bVec{p}_\gamma \times \bVec{p}_N +
{\cal F}_{15}\,\bVec{p}_\gamma \times \bVec{p}_Y +{\cal F}_{16}\,\bVec{p}_N
 \times 
\bVec{p}_Y + \right.\nonumber\\ && \left.
\bVec{p}_Y\cdot\bVec{p}_\gamma \times \bVec{p}_N 
\left({\cal F}_{18}\, \bVec{p}_\gamma +
{\cal F}_{19}\, \bVec{p}_N + {\cal F}_{20}\, \bVec{p}_Y\right)\right]
\end{eqnarray}
and
\begin{eqnarray} 
{\cal K}_{ij} &=& \delta_{ij}\,A + p_{\gamma ,i}\,{B}_j + 
p_{N,i}\,{C}_j + p_{Y,i}\,{D}_j  ~, ~~~~ i,j=x,y,z  ~~,
\end{eqnarray}
with 
\begin{eqnarray} 
A &=& -N \left[{\cal F}_1+{\cal F}_{14}\,\bVec{p}_N\cdot\bVec{p}_\gamma -
{\cal F}_{15}\,\bVec{p}_Y\cdot\bVec{p}_\gamma-{\cal F}_{16}\,\bVec{p}_N
\cdot\bVec{p}_Y\right]
~,\\
\bVec{B} &=& -N \left[({\cal F}_4-{\cal F}_{14}-{\cal F}_{19}\,
\bVec{p}_N\cdot\bVec{p}_Y)\,\bVec{p}_N 
+ ({\cal F}_{5}+{\cal F}_{15}-{\cal F}_{20}\,\bVec{p}_N\cdot\bVec{p}_Y)\,
\bVec{p}_Y 
\right] ~,\\
\bVec{C} &=& -N \left[({\cal F}_8+{\cal F}_{19}\,\bVec{p}_Y\cdot\bVec{p}
_\gamma)\,\bVec{p}_N 
+ ({\cal F}_{9}+{\cal F}_{16}+{\cal F}_{20}\,\bVec{p}_Y\cdot
\bVec{p}_\gamma)\,\bVec{p}_Y \right] ~,
\\
\bVec{D} &=& -N \left[({\cal F}_{12}+{\cal F}_{16}+{\cal F}_{19}\,\bVec{p}_N
\cdot\bVec{p}_\gamma)\,\bVec{p}_N 
+ ({\cal F}_{13}+{\cal F}_{20}\,\bVec{p}_N\cdot\bVec{p}_\gamma)\,\bVec{p}_Y
 \right]
~.
\end{eqnarray}
Furthermore, we have 
\begin{equation} 
N= 
\left(\frac{E_N+m_N}{2m_N}\right)^{\frac{1}{2}}
\left(\frac{E_Y+m_Y}{2m_Y}\right)^{\frac{1}{2}}
\sqrt{\frac{m_Y}{E_Y}}
\sqrt{\frac{m_N}{E_N}} ~.
\end{equation}

\section{Cross sections and Polarization observables}
\label{inclusive}

As shown in Ref.~\cite{yamamura99}, the cross section for the 
$\gamma +d \rightarrow  K^+ +Y+N$ process in a general frame 
is expressed as
\begin{eqnarray}
d\sigma (
\mu_Y \nu_Y \mu_N \nu_N,  \mu_d \, \epsilon 
    )
&=&
 \frac{(2\pi)^3}{4E_K E_\gamma}
\int\frac{d\bVec{p}_K}{(2\pi)^3}\frac{d\bVec{p}_Y}{(2\pi)^3}
\frac{d\bVec{p}_N}{(2\pi)^3} \nonumber \\
&&\times \left|\,\sqrt{2} \langle \,\Psi^{(-)}_{\bVec{q}_Y 
\mu_Y\, \nu_Y\, \mu_N\, \nu_N}
\,|\,t_{\gamma K}(1)
\,|\,\Psi_d\, \mu_d\,\rangle \,\right|^2 
\nonumber \\
&&\times (2\pi)^4 \delta^{(4)}
( P_d + Q - p_Y - p_N ) ~,
\label{ic1}
\end{eqnarray}
where the $\mu$'s and $\nu$'s denote spin and isospin  magnetic quantum
numbers and $\epsilon$ the photon polarization. The two-baryon final state
 state
$\Psi$ depends on $\bVec{q}_Y$, the (nonrelativistic) relative momentum of
 the
final hyperon and nucleon, while the momentum transfer to the
final state is denoted by  $\bVec{Q} = \bVec{p}_\gamma - \bVec{p}_K$.
The elementary operator $t_{\gamma K}(1)$ contains a label,
indicating that it acts only on one of 
the baryons. The kinematics and the elementary operator are kept in 
their relativistic
form, while nonrelativistic two-baryon wave functions are used.

In  Ref.~\cite{yamamura99}, we calculated cross sections in the total
momentum zero frame of
the final two baryons. However, the nuclear matrix element in
Eq.~(\ref{ic1}) is not invariant under Lorentz transformations.
The reason is the nonrelativistic treatment
of the two baryons in the deuteron and the final state.
Therefore, in the present paper we calculate all observables  in the
lab  frame (the deuteron rest frame) 
which corresponds to the  experimental situation.

The exclusive cross section is obtained by carrying out the integrations in 
Eq.~(\ref{ic1}) in the lab frame,

\begin{eqnarray}
\frac{d\sigma}{dp_K d\Omega_K d\Omega_Y}&=&
\frac{\bVec{p}^2_K}{(2\pi)^2\, 4E_\gamma E_K } 
\, \frac{ m_Y m_N \, |\bVec{p}_Y|}
{|E_N+E_Y-E_Y\frac{\bVec{Q}\cdot \bVec{p}_Y}{p_Y^2}| }
\nonumber \\
&&\times
\left|\,\sqrt{2}\,
\langle \,\Psi^{(-)}_{\bVec{q}_Y 
\mu_Y\nu_Y\mu_N\nu_N}\,|\,t_{\gamma K}(1)\,|\,
\Psi_d \mu_d\,\rangle \,\right|^2 ~,
\label{ic2}
\end{eqnarray}
where the relative momentum $\bVec{q}_Y$ is determined by 
the nonrelativistic relation
\begin{equation}
 \bVec{q}_Y = \bVec{p}_Y
            -\frac{m_Y}{m_Y+m_N}\bVec{Q}~.
\label{ic6}
\end{equation}

The  calculations  of the nuclear matrix element
are discussed in detail in Ref.~\cite{yamamura99},
but Eqs.~(3.13), (3.14) and (3.15) therein should be replaced 
by the expressions in the lab frame,

\begin{eqnarray}
\bVec{k}_Y &=& \bVec{q}_Y
            +\frac{m_Y}{m_Y+m_N}\bVec{Q}~,
\label{ic3}\\
 \bVec{k}_1 &=& \bVec{q}_Y
            -\frac{m_N}{m_Y+m_N}\bVec{Q}~,
\label{ic4}\\
 \bVec{q} &=& \bVec{q}_Y
            -\frac{m_N}{m_Y+m_N}\bVec{Q}~.
\label{ic5}
\end{eqnarray}

The inclusive $^2{\rm H}(\gamma,K^+)$ cross section
can be obtained by integrating Eq.~(\ref{ic2}) over 
$\Omega_Y$. However, this turns out to be inconvenient since
the integration limits
over $\Omega_Y$ are complicated and, secondly, the phase space
factor appearing in Eq.~(\ref{ic2}) can become singular.
We therefore take as the integration variable the  direction
of the hyperon momentum $\bVec{p}^{\rm cm}_Y$ in the c.m. frame
of the final two baryons. 

Thus, the inclusive cross section (in the lab frame) is expressed as

\begin{eqnarray}
\frac{d\sigma}{dp_K d\Omega_K}&=&
\frac{\bVec{p}^2_K}{(2\pi)^2\, 4E_\gamma E_K W } \sum_Y
 m_Y m_N \, |\bVec{p}^{\rm cm}_Y| \nonumber \\
&&\times \frac{1}{6}
 \sum_{\mu_d\, \epsilon}
 \sum_{\mu_Y\,\mu_N}
\sum_{ \nu_Y\,\nu_N}
\int d\hat{\bVec{p}}^{\rm cm}_Y \left|\,\sqrt{2}\,
\langle \,\Psi^{(-)}_{\bVec{q}_Y 
\mu_Y\nu_Y\mu_N\nu_N}\,|\,t_{\gamma K}(1)\,|\,
\Psi_d \mu_d\,\rangle \,\right|^2 ~,
\label{ic7}
\end{eqnarray}
where  $W^2=(P_d+Q)^2$.

In addition to the inclusive $^2{\rm H}(\gamma,K^+)$ and exclusive 
$^2{\rm H}(\gamma,K^+Y)$ cross sections,  we calculate 
the hyperon recoil polarization $P_y$, the
beam polarization asymmetry $\Sigma$
and the double polarizations $C_x$, $C_z$ which are
given by

\begin{eqnarray}
P_y &=& \frac{{\rm Tr}\{MM^+\sigma_y\}}
              {{\rm Tr}\{MM^+\}}, 
\label{ic8}\\
\Sigma &=& 
  \frac{{\rm Tr}\{M_{\epsilon =\epsilon_y} M^+_{\epsilon =\epsilon_y}\}
       -{\rm Tr}\{M_{\epsilon =\epsilon_x} M_{\epsilon =\epsilon_x}^+\}}
     {{\rm Tr}\{M_{\epsilon =\epsilon_y} M_{\epsilon =\epsilon_y}^+\}
     +{\rm Tr}\{M_{\epsilon =\epsilon_x} M_{\epsilon =\epsilon_x}^+\}},
\label{ic9}\\
C_x &=& \frac{{\rm Tr}\{M_{\epsilon =\epsilon_1}M_{\epsilon =\epsilon_1}
        ^+ \sigma_x\}}
    {{\rm Tr}\{M_{\epsilon =\epsilon_1}M_{\epsilon =\epsilon_1}^+\}}, 
\label{ic10}\\
C_z &=& \frac{{\rm Tr}\{M_{\epsilon =\epsilon_1}M_{\epsilon =\epsilon_1}
          ^+  \sigma_z\}}
     {{\rm Tr}\{M_{\epsilon =\epsilon_1}M_{\epsilon =\epsilon_1}^+\}}. 
\label{ic11}
\end{eqnarray}

In the definitions above, $M$ is the $K^+YN$ break-up  amplitude
\begin{equation}
M ( \mu_Y\, \mu_N\, ; \mu_d\, \epsilon )=
 \langle \,\Psi^{(-)}_{\bVec{q}_Y \mu_Y\, \nu_Y\, 
\mu_N\, \nu_N}
\,|\,t_{\gamma K}(1)
\,|\,\Psi_d\, \mu_d\,\rangle  
\label{ic12}
\end{equation}
with $M_{\epsilon =\epsilon_y}$ as 
the amplitude where the photon polarization 
points into the y-axis, $\bVec\epsilon =\bVec\epsilon_y$,
and so on.
The photon polarization
$\bVec\epsilon_1=-\frac{1}{\sqrt{2}}
(\bVec\epsilon_x+i\bVec\epsilon_y)$
describes the helicity state +1. 
The beam polarization asymmetry $\Sigma$ is obtained with
linearly polarized
photon, while the double polarization observables $C_x$ and $C_z$ are the
 hyperon 
polarization along with circularly polarized photons.

\newpage

\section{Results and discussions}
\label{results} 

As mentioned in Sec. I, the aim of this paper is twofold, i.e., to study
the $YN$ 
final-state interaction (FSI) effects as well as to extract the information 
on the elementary amplitude in the region where FSI effects are negligible.
For this purpose, we limit the outgoing kaon lab angles to the forward ones
less than $20^\circ$ because we expect other types of FSI would start to
show 
some effects for larger kaon angles. As a sample, the incoming photon energy of 
1.3 GeV is chosen throughout this paper.
We analyze observables for  two $YN$ forces, i.e., NSC89 \cite{nsc89} and 
NSC97f \cite{nsc97}, both of which give  the correct hypertriton binding
energy \cite{ours}. 
The NSC89 interaction is a soft-core one-boson-exchange model based on
Regge-pole theory, and a straightforward extension of the $NN$ model
through the application of SU(3).
The NSC97f model is an improvement of NSC89.
Reference \cite{nsc97} gives the six models from NSC97a through
NSC97f  classified 
by the different choices for the magnetic vector-meson 
$F/(F+D)$ ratio, which results in a different balance between
$^1S_0$ and $^3S_1$ $\Lambda N$ scattering lengths.  Among them only the
two models, 
NSC89 and NSC97f, reproduce the hypertriton binding energy, but still have
different 
$S$-wave scattering lengths \cite{ours,fewbody}. 
A rigorous four-body calculation \cite{nogga} 
has also shown that both of these two can not reproduce the bound states
of $^4_\Lambda$H and $^4_\Lambda$He.
Thus, the present analysis is expected to supply additional information 
on the $YN$ interaction via the continuum.

The deuteron wave function is generated by the Nijmegen93
 potential~\cite{nij93}. Some figures and discussions for the inclusive
 observables are
 similar to
those which are already given in the previous paper \cite{yamamura99}, but
we include them
for the convenience of the readers. Notice that we present all observables
in the deuteron rest frame as mentioned in Sec.\,III, which differs from the
former paper.

\subsection{Inclusive observables}

\begin{figure}[t]
\epsfxsize=8cm \epsfxsize=8cm
\centerline{\epsffile{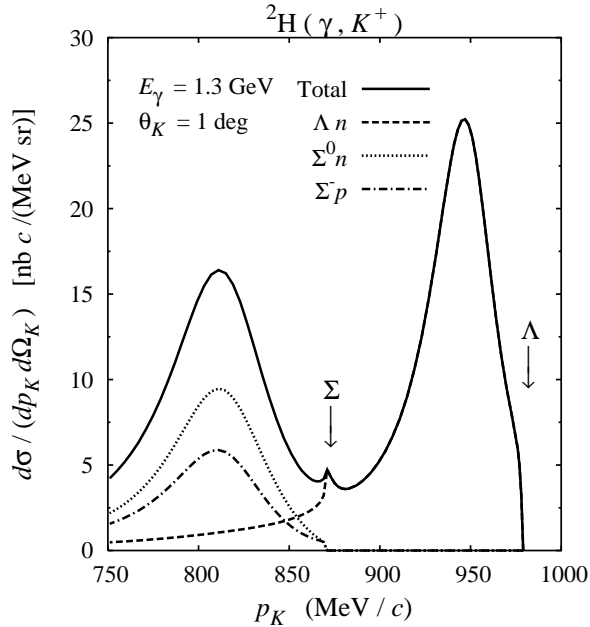}}
\caption{\label{fig1}
  Inclusive $^2{\rm H}(\gamma ,K^+)$ cross section as a function
  of kaon lab momentum $p_K$ for kaon lab angle $\theta_K=1^\circ$ and 
  photon beam energy $E_\gamma=1.3$ GeV. The NSC97f potential is used
  for the $YN$ sector. 
  The contributions from the $\Lambda n$,
  $\Sigma^0 n$ and $\Sigma^- p$ processes are shown separately
  and summed up.
  The arrows indicate the $\Lambda N$ and $\Sigma N$ thresholds.}
\end{figure}

\begin{figure}[htbp]
\epsfxsize=8cm
\centerline{\epsffile{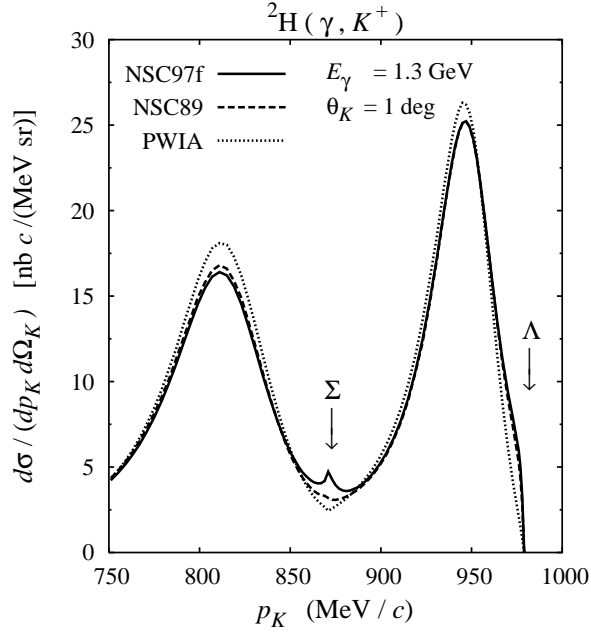}}
\caption{\label{fig2}
  Inclusive $^2{\rm H}(\gamma ,K^+)$ cross section as a function
  of kaon lab momentum $p_K$ for kaon lab angle
  $\theta_K=1^\circ$ and 
  photon beam energy $E_\gamma=1.3$ GeV.
  The PWIA result is compared to the two $YN$ force predictions. 
  The FSI effects are pronounced near the $\Lambda N$ 
  and $\Sigma N$ thresholds (indicated
  by the arrows) as well as on  top of the two peaks.}
\end{figure}

\begin{figure}[htbp]
\epsfxsize=10cm
\centerline{\epsffile{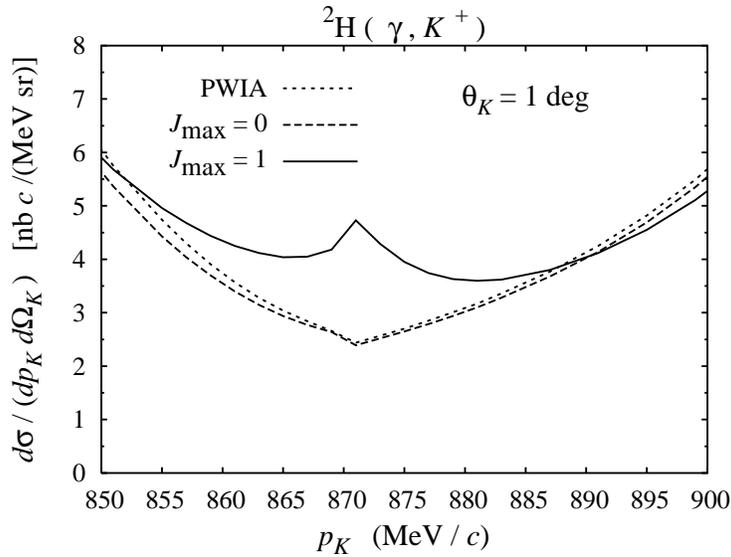}}
\caption{\label{fig3}
  The results in Fig.~\ref{fig2} for NSC97f and PWIA
  enlarged around the $\Sigma N$ threshold.
  The individual contributions
  of $\Lambda N$-$\Sigma N$ partial waves are also shown.}
\end{figure}

First, inclusive cross sections for the kaon lab angle $\theta_K=1^\circ$
are shown in Fig.~\ref{fig1} as a function of the  kaon lab momentum
$p_K$. 
The individual contributions of the $\Lambda n$, $\Sigma^0 n$ and
$\Sigma^- p$ processes are shown separately
and summed up. The $YN$ final-state interaction NSC97f is used. 
The two pronounced peaks are seen around $p_K=950$ and $p_K=810$ MeV/$c$.
These $p_K$ values  correspond to the quasifree scattering (QFS)
 condition 
where one of the nucleons in the deuteron sits as a spectator and has zero
 momentum.
In Fig.~\ref{fig2}, the inclusive cross sections are compared with the
result obtained by using the plane wave impulse approximation (PWIA) and
the result
generated by the other $YN$ interaction NSC89.
The two peaks are dominated by the PWIA contribution but some FSI 
effects can be seen especially around $p_K=810$ MeV/$c$, about which we will
also discuss in the next subsection.  
In Fig.~\ref{fig2} we also see the FSI effects around the $\Lambda N$ and
the $\Sigma N$ thresholds. Around the $\Sigma N$ threshold, the NSC97f result 
shows  a prominent cusp-like structure, whereas the NSC89 one displays only a
 little 
deviation from the PWIA result. This fact can be traced back to the location
 of 
the $S$-matrix pole for the $\Lambda N$-$\Sigma N$ system around the
$\Sigma N$ threshold, the full detail of which has been discussed in 
Refs.~\cite{yamamura99} and \cite{tpole}. The structure produced by NSC97f 
with different $\Lambda N$-$\Sigma N$ partial wave contributions is enlarged
in Fig.~\ref{fig3}.  
The line indicated by $J_{\rm max}=0$ incorporates the $^1S_0$ and $^3P_0$
components
and $J_{\rm max}=1$ includes all the partial waves up to $J=1$.  
It shows that the $J=1$ state is responsible for this
structure,
which is consistent with the known fact that the $S$-matrix pole mentioned
above  exists in
the $^3S_1$-$^3D_1$ state.

\begin{figure}[t]
\epsfxsize=8cm
\centerline{\epsffile{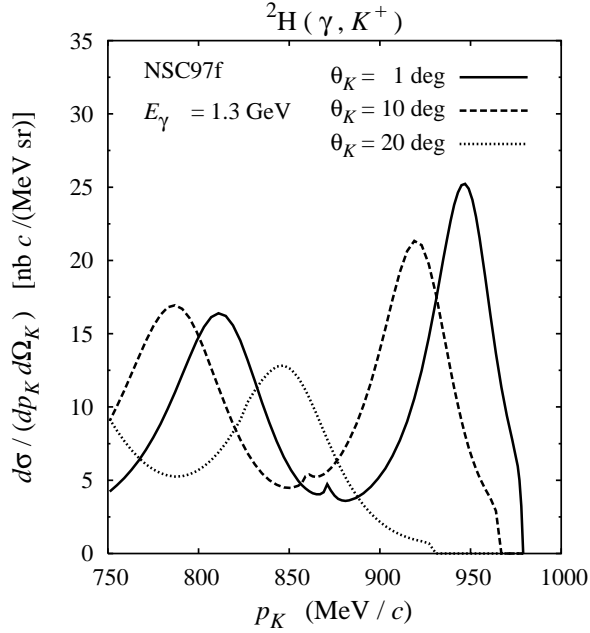}}
\caption{\label{fig4}
  Inclusive $^2{\rm H}(\gamma ,K^+)$ cross section  as a function
  of the kaon lab momentum $p_K$ for kaon lab angles
  $\theta_K=1^\circ, 
  10^\circ$ and $20^\circ$ and 
  photon beam energy $E_\gamma=1.3$ GeV.
  The $YN$ interaction NSC97f has been used to obtain these results. }
\end{figure}

In Fig.~\ref{fig4}, the inclusive cross sections for three 
different kaon lab angles, i.e., $\theta_K=1^\circ, 10^\circ$ and $20^\circ$
are shown,
where the cross section maxima shift as the kaon angle increases.
We confirmed that, in contrast to the $\theta_K=1^\circ$
case, the cross sections at $\theta_K=20^\circ$ have
little $YN$ FSI effects in the region of $p_K$ larger than $750$ MeV/$c$.
Also we verified that the
$J_{\rm max}=1$ results are converged with regard to the angular momentum
decomposition of the $YN$ system throughout the figures for the inclusive 
cross sections.

\begin{figure}[htbp]
\epsfxsize=10cm
\centerline{\epsffile{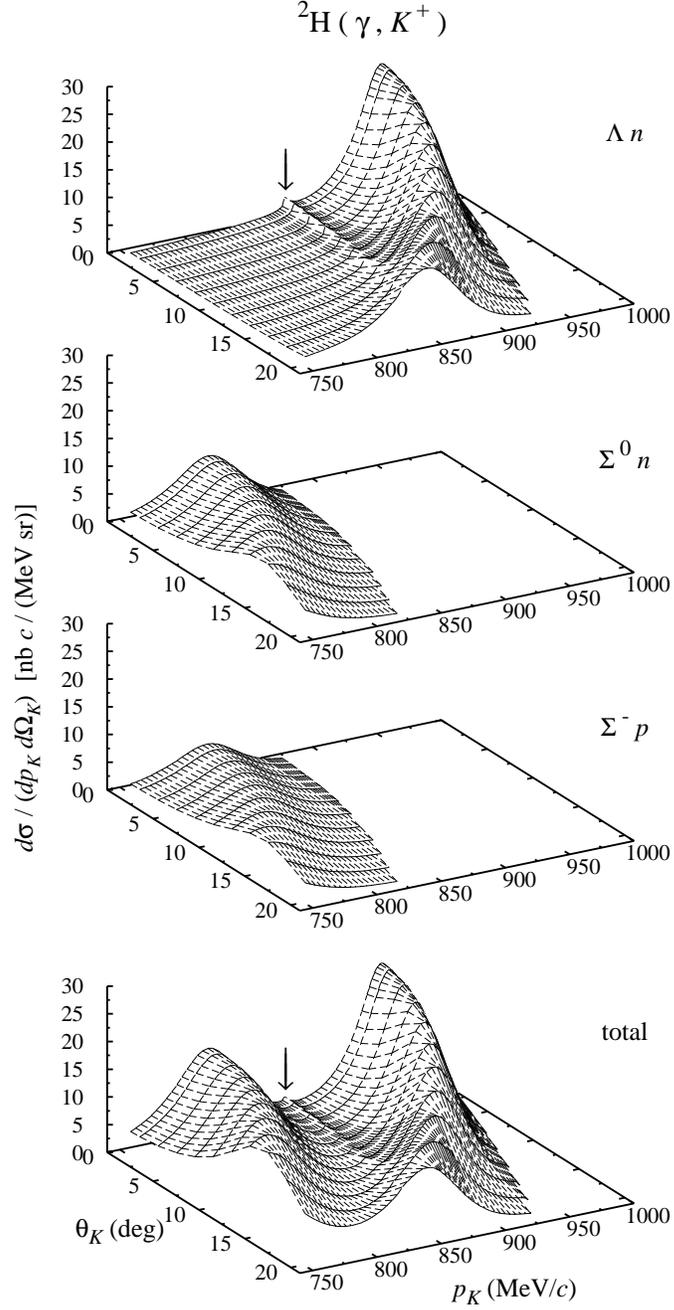}}
\caption{\label{fig6}
  Inclusive $^2{\rm H}(\gamma ,K^+)$ cross section as a function
  of kaon lab momentum $p_K$ and angle $\theta_K$ 
  for photon beam energy
  $E_\gamma=1.3$ GeV. The NSC97f $YN$ interaction is used. 
  The contributions from the $\Lambda n$,
  $\Sigma^0 n$ and $\Sigma^- p$ processes are shown separately
  and summed up.
  The arrows indicate the $\Sigma N$ threshold.}
\end{figure}

Finally, in Fig.~\ref{fig6} we give the three-dimensional plots of the
 inclusive 
cross sections
as a function of lab kaon momentum $p_K$ and angle $\theta_K$, where
individual contributions of the $\Lambda n$, $\Sigma^0 n$ and
$\Sigma^- p$ processes are also shown.
On the plane of $p_K=750\sim 1000$ MeV/$c$ and $\theta_K=1\sim 20^\circ$,
we can identify the area where the cross sections are large. 
In particular, the two peaks seen in Fig.~\ref{fig1} form two ridges
in the $p_K-\theta_K$ plane, and it is confirmed that these 
ridges run along the $p_K-\theta_K$ values which satisfy the
QFS condition. 

\clearpage


\newpage

\subsection{Exclusive observables}

\begin{figure}[htbp]
\epsfxsize=8cm
\centerline{\epsffile{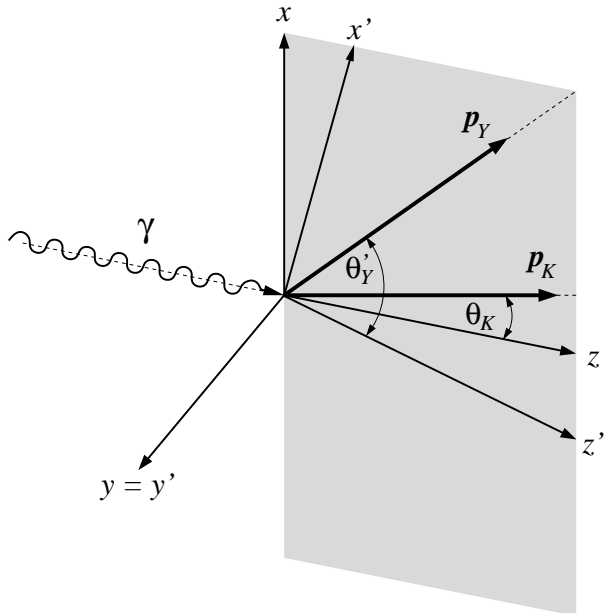}}
\caption{\label{figaxis}
  Kinematics in the deuteron rest frame. The $z$ axis points 
 into  the photon beam direction $\bVec p_{\gamma}$ and the kaon lies
  in the $x$-$z$ plane. The momentum transferred to the
  $YN$ system, $\bVec p_{\gamma}-\bVec p_K$, defines
  the $z'$ axis. The hyperon angle $\theta_Y'$ is measured from
  the $z'$ axis.}
\end{figure}   

There are many ways to present exclusive observables including the
 polarization 
ones. Here, we show five observables, i.e., differential cross section 
${d\sigma}/{dp_K d\Omega_K d\Omega_Y}$, 
hyperon polarization
$P_y$, double polarizations $C_z$ and $C_x$, and the beam polarization
asymmetry
$\Sigma$, as a function of the hyperon angle for fixed kaon momentum 
and kaon angle.  
As kaon momenta, we select $p_K=975$ MeV/$c$ (just above the $\Lambda n$
threshold), $944$ MeV/$c$ (close to  the $\Lambda$ QFS peak),
 $870$ and $860$ MeV/$c$ (near the $\Sigma N$ threshold),  and
$810$ MeV/$c$ (on the $\Sigma$ QFS peak). 

In Fig.~\ref{figaxis}, an illustration of the kinematics is given.
The $z$ axis points into the photon beam direction and the kaon lies 
in the $x$-$z$ plane, as usual. However, for the outgoing $YN$
system, Li and Wright \cite{li91} found a characteristic axis, which
corresponds to the direction of the momentum transfer
 $\bVec p_{\gamma}-\bVec p_K$
 to the $YN$ system
and defines the $z'$ axis as shown in Fig.~\ref{figaxis}. Thus a new
 coordinate
system $x'$-$y$-$z'$ is introduced with the $x'$ axis defined by
 $\hat{\bVec{e}}_y \times\hat{\bVec{e}}_{z'}$. 
We use  this primed coordinate
system as the reference system for the hyperon angles, which are denoted by 
$\theta_Y'$ and $\phi_Y'$. The hyperon
polarization is defined with respect to the ordinary $x$-$y$-$z$ system.
For all results given below, we take  $\phi_Y'=0$ and 
the hyperon is on the $x'$-$z'$ (and $x$-$z$) plane.


\begin{figure}[t]
\epsfxsize=10cm
\centerline{\epsffile{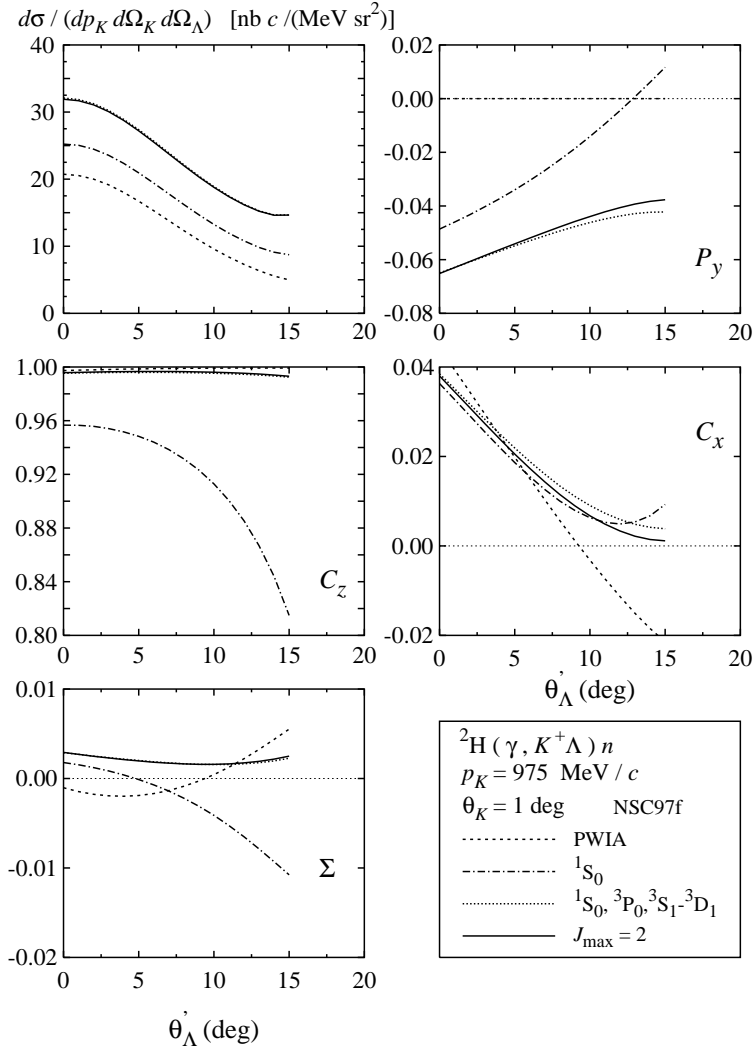}}
\caption{\label{fig21}
  Differential cross section and polarization observables 
  $P_y$, $C_z$, $C_x$ and $\Sigma$, 
  for the $^2{\rm H}(\gamma ,K^+ \Lambda)n$ process as a function
  of the hyperon lab angle $\theta_{\Lambda}'$.
  The kaon lab momentum and angle are fixed
  at $p_k=975$ MeV/$c$ and $\theta_K=1^\circ$.
  The $YN$ interaction NSC97f is used and
  different partial wave contributions are shown.
  The possible $\Lambda$ angles are limited to  
  less than  $17^\circ$.}
\end{figure}   

\begin{figure}[htbp]
\epsfxsize=10cm
\centerline{\epsffile{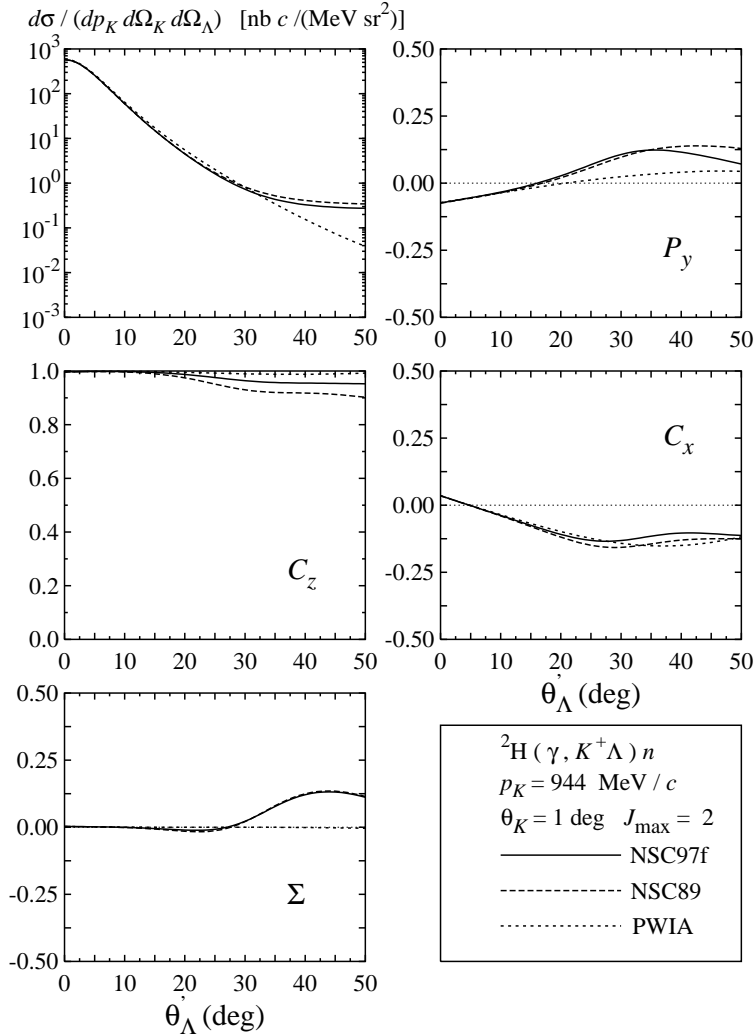}}
\caption{\label{fig19}
  Same as Fig.~\ref{fig21}, but for kaon lab momentum $p_k=944$ MeV/$c$. 
The results 
  obtained from NSC97f and NSC89 are compared with the one from PWIA.}
\end{figure}   


\begin{figure}[htbp]
\epsfxsize=10cm
\centerline{\epsffile{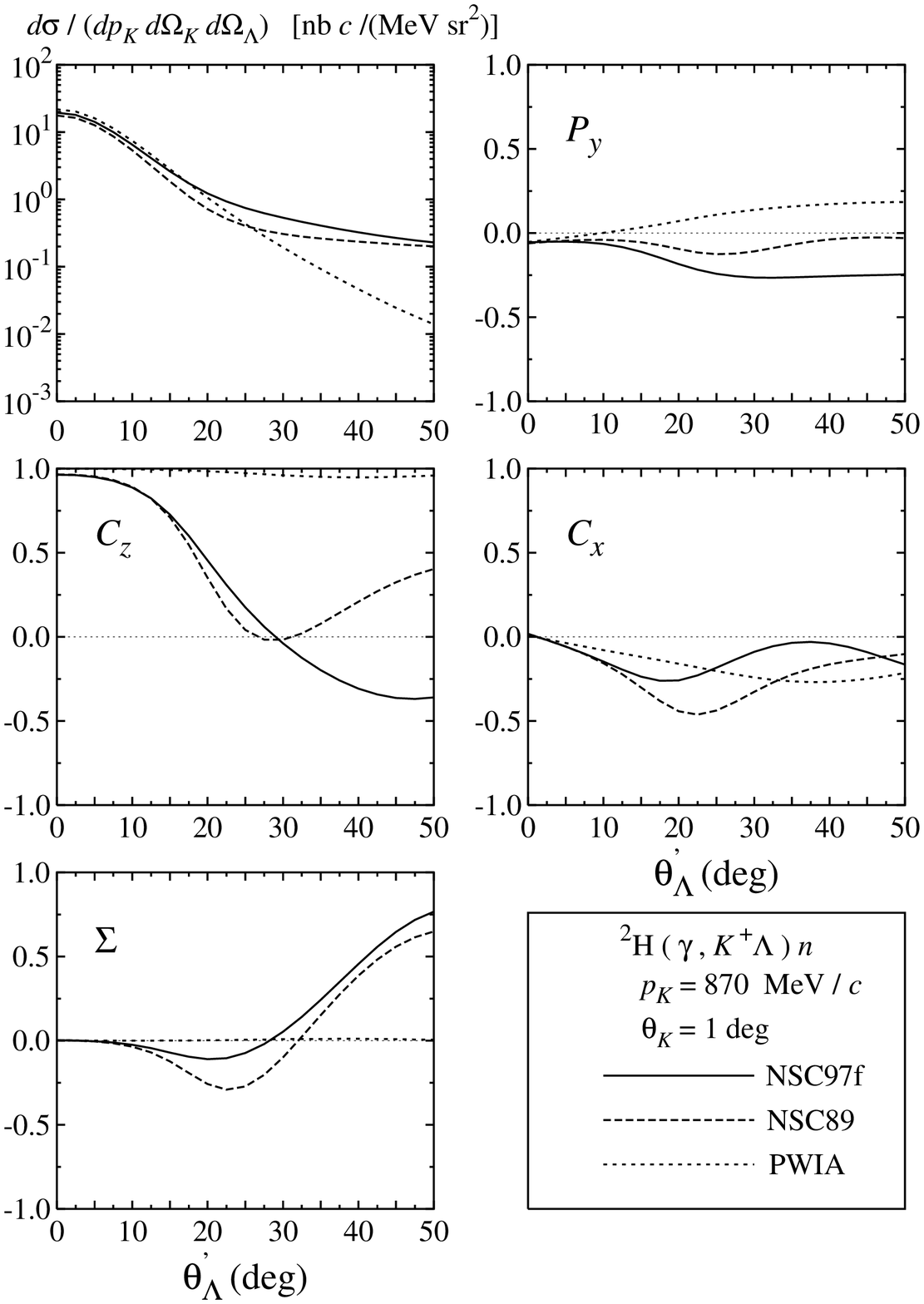}}
\caption{\label{fig10}
  Differential cross section and polarization observables $P_y$, $C_z$, $C_x$
  and $\Sigma$ 
  for the $^2{\rm H}(\gamma ,K^+ \Lambda)n$ process as functions of $\Lambda$ angle 
  $\theta_{\Lambda}'$. The kaon lab momentum and angle are fixed at
  $p_k=870$ MeV/$c$ 
  and $\theta_K=1^\circ$. The results by the $YN$ interactions NSC97f and
   NSC89
  are compared with that by PWIA.}
\end{figure}   

\begin{figure}[htbp]
\epsfxsize=10cm
\centerline{\epsffile{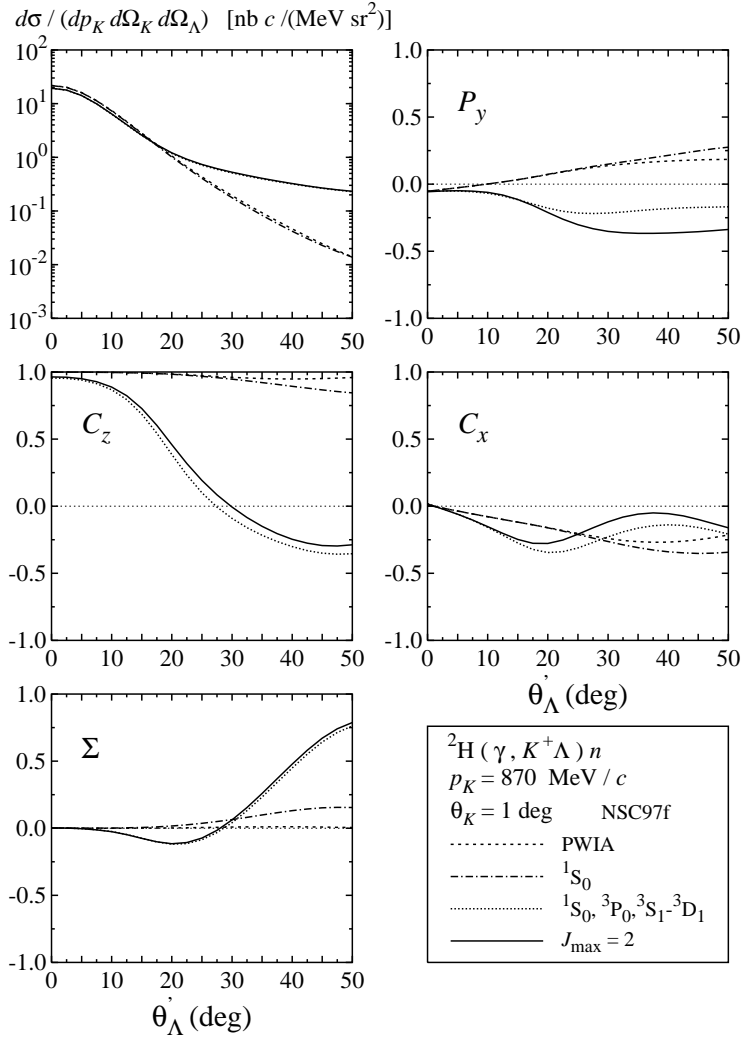}}
\caption{\label{fig16}
  Same as Fig.~\ref{fig10}, but the $YN$ interaction NSC97f is used and
  different partial wave contributions are shown.}
\end{figure}   


\begin{figure}[htbp]
\epsfxsize=10cm
\centerline{\epsffile{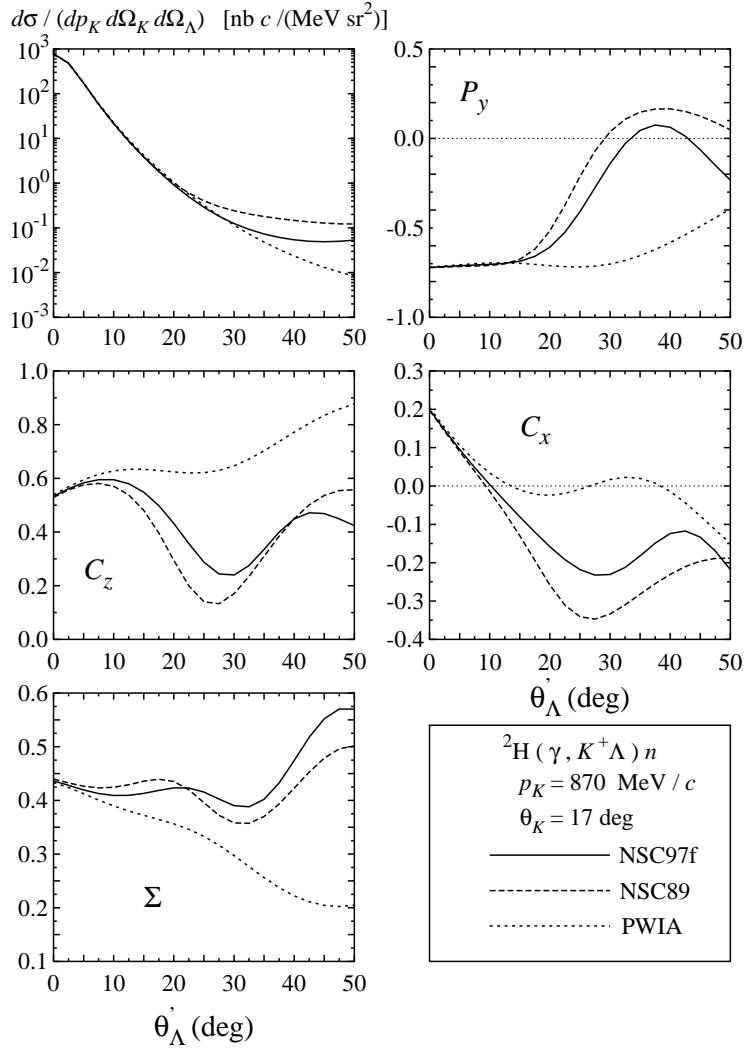}}
\caption{\label{fig11}
  Same as Fig.~\ref{fig10}, but for kaon lab angle $\theta_K=17^\circ$.}
\end{figure}   


\begin{figure}[htbp]
\epsfxsize=10cm
\centerline{\epsffile{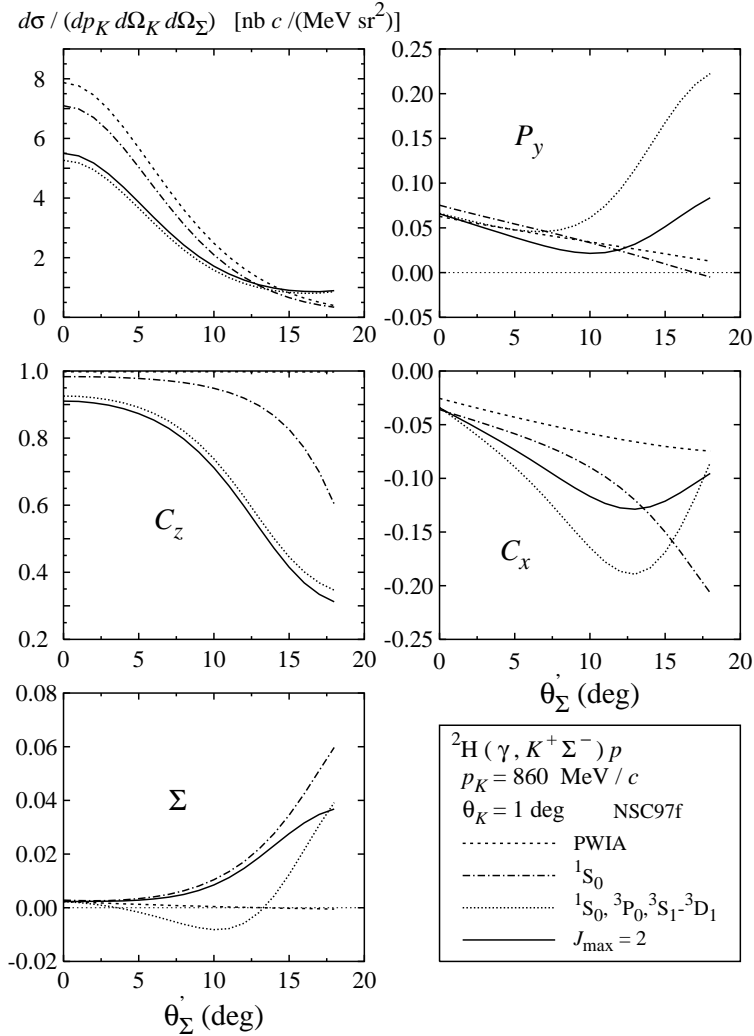}}
\caption{\label{fig18x}
  Differential cross section and polarization observables 
  $P_y$, $C_z$, $C_x$ and $\Sigma$ 
  for the $^2{\rm H}(\gamma ,K^+ \Sigma^-)p$ process
  as functions
  of the $\Sigma$ angle $\theta_{\Sigma}'$.
  The kaon lab momentum and angle are fixed
  at $p_k=860$ MeV/$c$ and $\theta_K=1^\circ$.
  The $YN$ interaction NSC97f is used and
  different partial wave contributions are shown.
  The available $\Sigma$  lab  angles are limited to 
  less than  $\theta_{\Sigma}'=21^\circ$.}
\end{figure}   

\begin{figure}[htbp]
\epsfxsize=10cm
\centerline{\epsffile{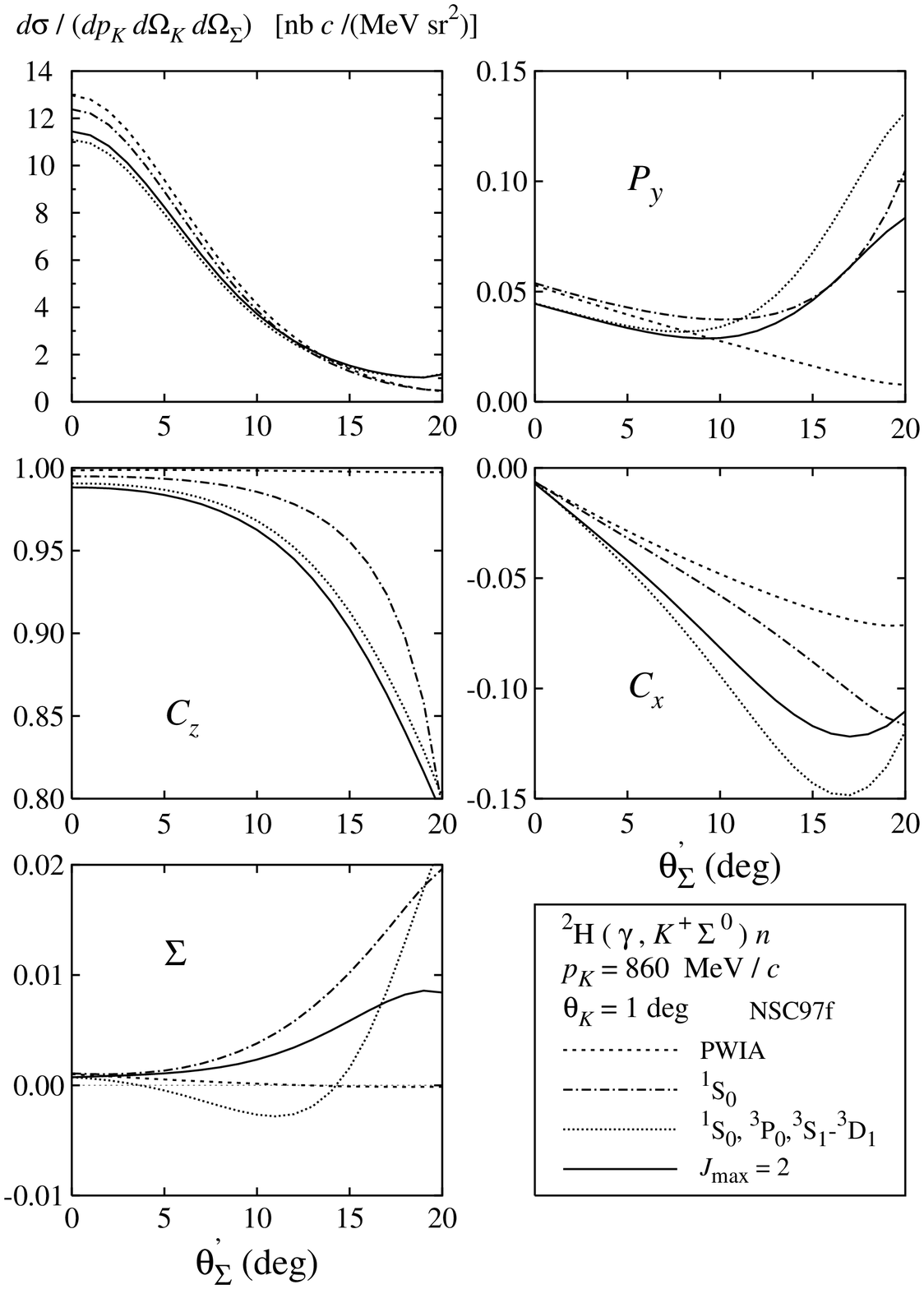}}
\caption{\label{fig22}
Same as Fig.~\ref{fig18x}, but for the $^2{\rm H}(\gamma ,K^+ \Sigma^0)n$ process.}
\end{figure}   

\begin{figure}[htbp]
\epsfxsize=10cm
\centerline{\epsffile{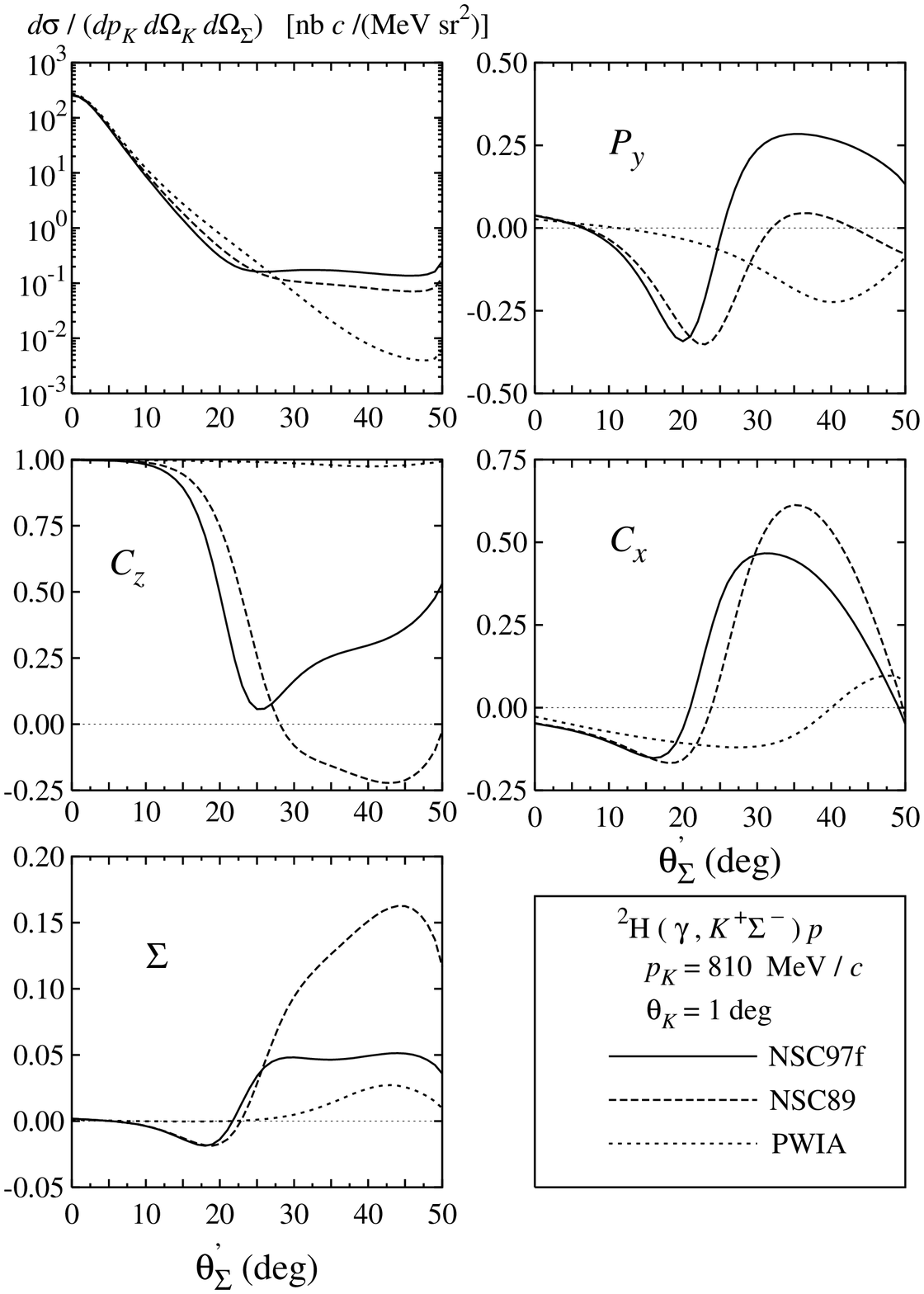}}
\caption{\label{fig12}
  Differential cross section and polarization observables 
  $P_y$, $C_z$, $C_x$ and $\Sigma$ for the $^2{\rm H}(\gamma ,K^+ \Sigma^-)p$ process
  as functions of the  $\Sigma$ angle $\theta_{\Sigma}'$.
  The kaon lab momentum and angle are fixed
  at $p_k=810$ MeV/$c$ and $\theta_K=1^\circ$.
  The results by the $YN$ interactions NSC97f and NSC89
  are compared with that by PWIA.  
  }
\end{figure}   

\begin{figure}[htbp]
\epsfxsize=10cm
\centerline{\epsffile{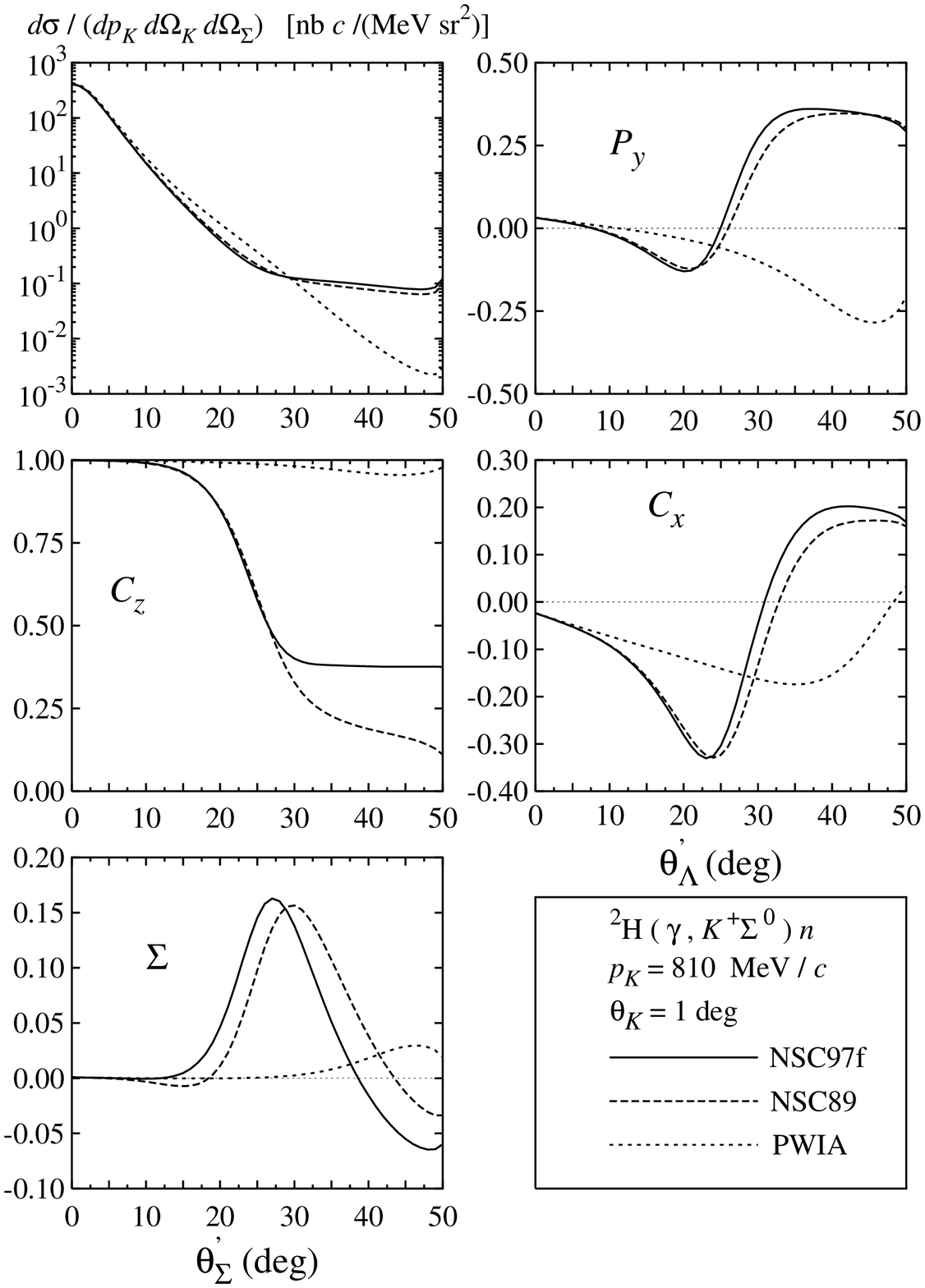}}
\caption{\label{fig14}
  Same as Fig.~\ref{fig12}, but for the $^2{\rm H}(\gamma ,K^+ \Sigma^0)n$ process.}
\end{figure}   



\begin{figure}[htbp]
\epsfxsize=10cm
\centerline{\epsffile{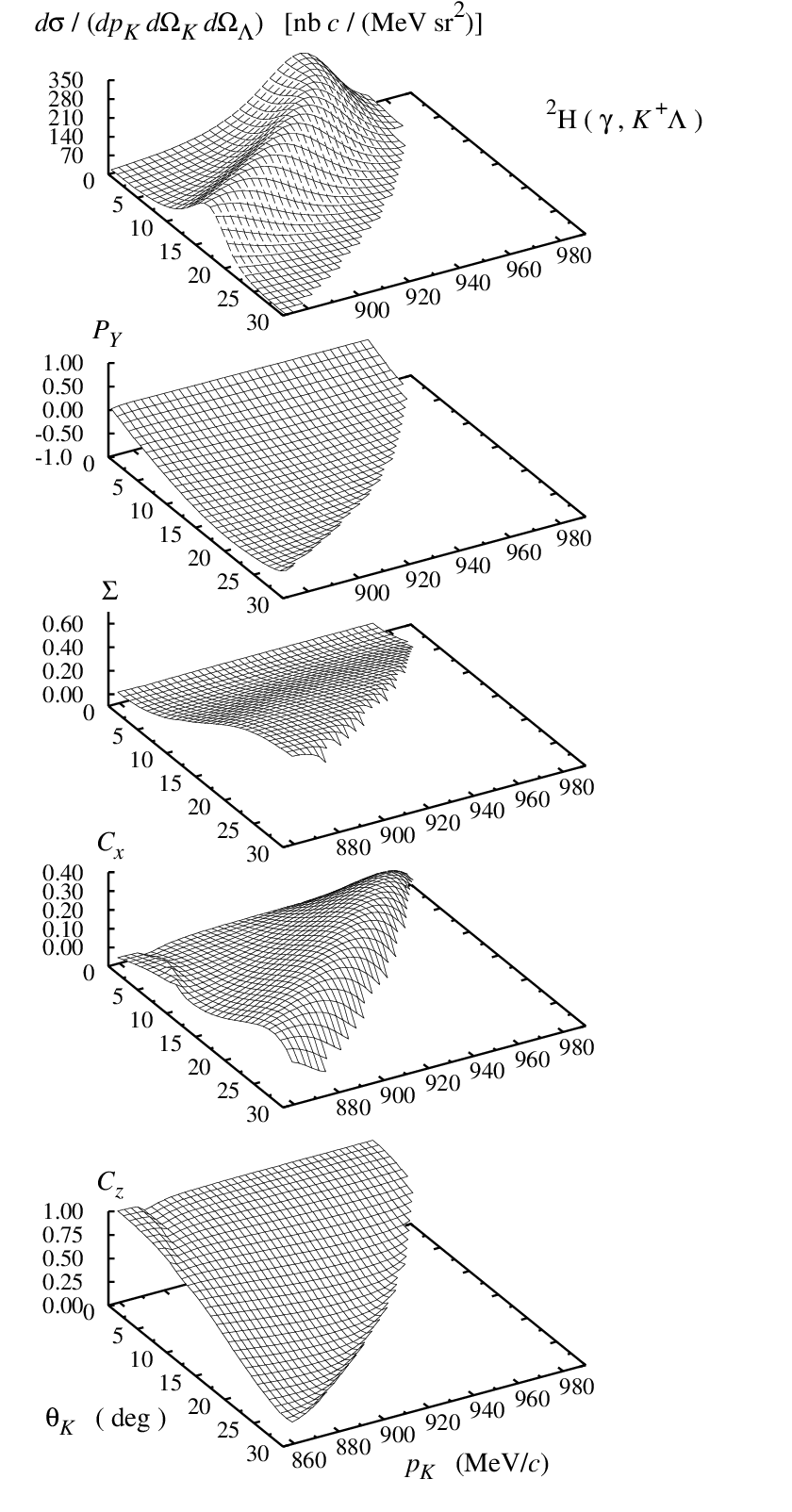}}
\caption{\label{fig7}
  Three-dimensional plots of exclusive observables
  ${d\sigma}/{dp_K d\Omega_K d\Omega_{\Lambda}}$,
  $P_y$, $\Sigma$, $C_x$ and $C_z$.
  for the $^2{\rm H}(\gamma ,K^+ \Lambda)n$ process
  as functions of kaon lab momentum
  $p_k$ and lab angle $\theta_K$.
  The $\Lambda$ lab angle is fixed at
  $\theta_{\Lambda}'=5^\circ$.}
\end{figure}   





Now, in Fig.~\ref{fig21}  the five observables 
in the $^2{\rm H}(\gamma ,K^+ \Lambda)n$ process
are shown as a function of  $\theta_{\Lambda}'$
for $p_k=975$ MeV/$c$ just above 
the $\Lambda n$ threshold. Here PWIA is compared with
the FSI results obtained by the NSC97f $YN$ interaction. The dominant
$YN$ partial wave contributions are also shown.
The line indicated by $J_{\rm max}=2$ incorporates all the partial wave
up to $J=2$, and gives the converged results.  
The differential cross sections clearly indicate
FSI effects starting from $\theta_{\Lambda}'=0^\circ$  to which 
$^1S_0$ and $^3S_1$-$^3D_1$ force components contribute.
In the double 
polarization $C_z$, $^1S_0$ shows a visible effect but
the inclusion of the $^3S_1$ cancels it and shifts it on the
same line as PWIA.
Please note, for the sake of better visibility of the effects we change 
the scales in various panels in Fig.~\ref{fig21} and the following ones.

Figure~\ref{fig19} shows the results for 
$p_K=944$ MeV/$c$ which is close to one of the  peak positions
in Fig.~\ref{fig2} (inclusive cross section).
The FSI effects for NSC97f and NSC89 are compared with
PWIA.
In the forward peak  of the differential cross sections, PWIA
dominates up to $\theta_{\Lambda}'=30^\circ$ which forms the dominant
part of the
$\Lambda$ QFS
peak in Fig.~\ref{fig2}.
(To be precise, the QFS condition corresponds to $\theta_K=1.6^\circ$
and $\theta'_\Lambda=0$ at this $p_K$ value.)
However, FSI  still shows some effects above 
$\theta_{\Lambda}'=30^\circ$.
The NSC97f and NSC89 interactions give almost the same effects
except for $C_z$.
  
Next, the results of $^2{\rm H}(\gamma ,K^+ \Lambda)n$  for $p_K$=870 MeV/$c$ 
and $\theta_K=1^\circ$ 
which is very close to the $\Sigma N$ threshold
are shown 
 in Figs.~\ref{fig10} and \ref{fig16}.
Figure~\ref{fig16} exhibits the dominant
partial wave contributions in NSC97f.
Because of the closeness to the threshold, the differential
cross sections show a rather flat behavior
and FSI effects are significant. In particular, the  double
polarization $C_z$ with FSI deviates from PWIA even at 
$\theta_{\Lambda}'=0^\circ$ and drastically above  
$\theta_{\Lambda}'=20^\circ$. Also NSC97f and NSC89 
are clearly distinct from each other in this case.
Figure~\ref{fig16} reveals the fact that these effects 
come from $\Lambda N$-$\Sigma N$ $^3S_1$-$^3D_1$ component
which is expected from the $t$-matrix pole existence in this partial
wave
as mentioned in the previous subsection.
 
We also show the results for $p_K$=870 MeV/$c$ but for a
larger kaon angle $\theta_K=17^\circ$ 
in Fig.~\ref{fig11}.
This case is rather far from the $\Sigma N$ threshold, and 
shows the forward-peak cross sections, but 
prominent FSI effects are still seen in  $P_y$ above 
$\theta_{\Lambda}'=20^\circ$. 

Now, let us give the results for the $\Sigma^- p$ and
$\Sigma^0 n$ exit channels. 
Figures~\ref{fig18x} ($\Sigma^- p$) and \ref{fig22} ($\Sigma^0 n$)
show the observables for
$p_K=860$ MeV/$c$ and $\theta_K=1^\circ$, which is still close
to the $\Sigma N$ threshold, and the available lab angles of
the hyperon $\theta_{\Sigma}'$ are limited to less than $21^\circ$.
As in Fig.~\ref{fig16}, FSI effects are prominent.
The $\Sigma^- p$ channel gives more visible FSI effects
than $\Sigma^0 n$, where both of the $\frac{1}{2}$ and $\frac{3}{2}$
$\Sigma N$ isospin components are included.
Not only the $^3S_1$-$^3D_1$, but also $^1S_0$
contributions can be seen,
especially in $C_z$.

The results for $p_K=810$ MeV/$c$ and $\theta_K=1^\circ$
which corresponds to the $\Sigma$ QFS peak position in Fig.~\ref{fig2}
are given in Fig.~\ref{fig12} ($\Sigma^- p$ channel)
and in Fig.~\ref{fig14} ($\Sigma^0 n$ channel).
In contrast to the case for $p_K=944$ MeV/$c$ ($\Lambda$ QFS),
most observables show significant FSI effects starting
from  $\theta_{\Sigma}'$ less than $20^\circ$. This is the
reason why fairly large FSI effects are seen on the top of 
the $\Sigma$ QFS peak around $p_K=810$ MeV/$c$ in Fig.~\ref{fig2}.  

In Fig.~\ref{fig7},
the three-dimensional plots of the five observables
as functions of kaon lab momentum $p_k$ and angle $\theta_K$
are given
for the $^2{\rm H}(\gamma ,K^+ \Lambda)n$ as an example.
The hyperon lab angle is fixed at the forward angle 
$\theta_Y'=5^\circ$. As in the inclusive case (Fig.~\ref{fig6})
we give the information on the $\theta_K$ dependence of the 
observables over the range of $\theta_K$ less than $30^\circ$. 

Before closing this section, we would like to discuss the influence
from other rescattering processes which are not included in the 
present analysis.
As mentioned in Sec.\,I, Refs.~\cite{maxwell2004} and \cite{salam2004}
have evaluated some of those rescattering contributions. The former
incorporates
several two-body kaon production mechanisms, 
and the latter includes
the pion mediated process $\gamma d\to \pi NN\to KYN$
in addition to the $YN$ and $KN$ rescattering.
In considering the effects of these rescattering processes,
we stress that the present analysis focuses on a restricted
kinematical region where 
the  kaon is scattered forward with large momentum
and small relative energy is distributed between
the hyperon and the nucleon.

In Fig.~\ref{fig19} ($p_K=944$ MeV/$c$, $\theta_K=1^\circ$), 
$T_K$(kinetic energy of the kaon) is $572$ MeV, 
but 
$T_{\Lambda-N}$(relative energy between $\Lambda$ and $N$)
is $25$ MeV. The QFS condition further determines the $\Lambda$
scattering angle $\theta_{\Lambda}'=0$ ($\theta_{\Lambda}=1.6^\circ$)
where $T_\Lambda$($\Lambda$ kinetic energy)$=56$ MeV and
$T_N=0$.   
The extreme cases are given by Figs.~\ref{fig18x} and \ref{fig22}, 
which are close to the $\Sigma N$ threshold and
$T_{\Sigma-N}$ is only $7$ MeV, while $T_K=498$ MeV.

In contrast to this, Ref.~\cite{maxwell2004} supposes a 
large relative energy between $Y$ and $N$ in the final state,
for example, Fig.~7 of Ref.~\cite{maxwell2004} shows
the case of $E_\gamma=1500$ MeV, $\theta_K=15^\circ$,
and $T_K=500$ (or 200) MeV, where
$T_{\Lambda-N}$ amounts to 221 (419) MeV.
Thus the $YN$ relative energies in the present work are at least 
10 times smaller than in Ref.~\cite{maxwell2004}.


Reference \cite{salam2004} focuses on the 
effects of the pion mediated process,
giving the total cross sections
and the semi-inclusive results 
where the contributions from various $p_K$ are integrated.
Recently, the same method has been applied in
Ref.~\cite{osaka04}
to the exclusive cross sections 
for just the same final
states as in  Fig.~\ref{fig10} ($p_K$=870 MeV/$c$), as well as 
Figs.~\ref{fig12} and \ref{fig14} ($p_K=810$MeV/$c$).
Important is  to note that
no visible effect of the pion mediated process
is seen.
The effects of
this process appear at small
$p_K$ values such as 400 MeV/$c$ in the inclusive cross sections
for $E_\gamma=1140$ MeV.  

From all the facts mentioned above, we conclude that in the
kinematical
region where the present
analysis concentrates, either QFS or $YN$ FSI is expected.

\clearpage

\section{Summary and Conclusion}
\label{conclusion}

We have evaluated single and double polarization observables
as well as the inclusive and exclusive cross sections for the
$\gamma +d \rightarrow  K^+ +Y+N$  reactions at $E_\gamma=1.3$ GeV,
which are immediately accessible experimentally by electron beam 
facilities such as JLab and SPring8. 
An updated photoproduction operator
for the $\gamma +N \rightarrow  K^+ +Y$ process and modern
hyperon-nucleon forces, NSC89 and NSC97f are used.
The kinematical region of the kaon scattered forward with 
large momentum is thoroughly investigated where either the quasifree
scattering
or the $YN$ final-state interaction is expected.
The quasifree scattering is found to generate large cross
sections and to
form the ridges on the $p_K-\theta_K$ plane where one could extract
the information on the elementary kaon photoproduction processes.
The $YN$ final-state interaction effects are significant,
especially around the $\Lambda N$ and $\Sigma N$ thresholds.
Around the $\Lambda N$ thresholds, both of the
$^1S_0$ and $^3S_1$
$YN$ force components show the effects, while
around the $\Sigma N$ thresholds, $^3S_1$-$^3D_1$ gives the main
contribution. In the latter case
the two $YN$ forces give quite different predictions, reflecting 
the different structure.
The polarization observables, in particular the double polarization
$C_z$,  are sensitive to
the final-state interaction effects.
Precise data would help to clarify the property of 
the $\Lambda N$-$\Sigma N$ interaction and also
help to extract the elementary
amplitude from the quasifree scattering region.

\acknowledgements
K.M. thanks the few-body group of the Ruhr 
University Bochum for the kind hospitality during his stay. 
T.M. thanks the Okayama University of Science for the hospitality
extended to him during his stay in Okayama and the support from
the Faculty of Mathematics and Sciences as well as from the
Hibah Pascasarjana.
This work was supported by "Academic Frontier"
Project for Private Universities: matching fund
subsidy from MEXT (Ministry of Education, Culture, Sports,
 Science and Technology), Japan.


\end{document}